\documentstyle[preprint,aps,epsfig]{revtex}

\begin{document}
\draft

\title{Roles of proton-neutron interactions in \\
 alpha-like four-nucleon correlations}

\author{M. Hasegawa and K. Kaneko}
\address{
Laboratory of Physics, Fukuoka Dental College, Fukuoka 814-0193, Japan}
\address{
Department of Physics, Kyushu Sangyo University, Fukuoka 813-8503,
 Japan}
\date{\today}
\maketitle

\begin{abstract}
  An extended pairing plus $QQ$ force model, which has been shown to
 successfully explain the nuclear binding energy and related quantities
 such as the symmetry energy, is applied to study the alpha-like
 four-nucleon correlations in $1f_{7/2}$ shell nuclei.
 The double difference of binding energies, which displays a
 characteristic behavior at $N \approx Z$, is interpreted in terms of
 the alpha-like correlations.  Important roles of proton-neutron
 interactions forming the alpha-like correlated structure are
 discussed.
 \end{abstract}
 

\narrowtext


  Recently, the present authors \cite{Hase1,Kane} have proposed
 a schematic interaction for $N \approx Z$ nuclei in $1f_{7/2}$ 
 and $1g_{9/2}$ shell regions and have applied it to explain double
 differences of binding energies discussed in Refs.
 \cite{Zhang,Brenner}. The double difference of binding energies
 of neighboring even-even nuclei was discussed to investigate
 $\alpha$-like four-nucleon correlations in early work
 \cite{Dussel,Gambhir}.  The calculations in Refs. \cite{Hase1,Kane}
 showed that the schematic interaction is applicable to the
 $1f_{7/2}$ shell nuclei where the $\alpha$-like four-nucleon
 correlations are very important
 \cite{Michel,Wada,Ohkubo2,Yamaya,Hase2,Suppl}.
 Our interaction, which is composed of the isospin-invariant
 monopole pairing ($P_0$), quadrupole pairing ($P_2$),
 quadrupole-quadrupole ($QQ$) forces and $J$-independent isoscalar
 proton-neutron ({\it p-n}) force ($V^{\tau=0}_{\pi \nu}$),
 is useful to study the roles of interactions.
   The purpose of this note is to investigate cooperation of the
 $P_0+P_2+QQ$ force and $V^{\tau=0}_{\pi \nu}$, or competition of
 the {\it p-n} and like-nucleon ({\it p-p, n-n}) interactions, in the
 $\alpha$-like four-nucleon correlations.

   It is well known that the single $j$ space $(f_{7/2})^n$ is a
 reasonable model space for the $1f_{7/2}$ shell nuclei \cite{MBZ}.
 We adopted the $(f_{7/2})^n$ model with the
 $P_0+P_2+QQ+V^{\tau=0}_{\pi \nu}$ interaction in Refs.
 \cite{Hase1,Kane}.
 Our effective Hamiltonian which describes valence nucleons outisde
 the doubly-closed-shell core $^{40}$Ca is given by
\begin{eqnarray}
 H &=& \epsilon_{7/2} \, {\hat n} + V_{PQ}
     + V^{\tau =0}_{\pi \nu}, \label{eq:1} \\
 V_{PQ} &=& V(P_0) +V(P_2) +V(QQ)   \nonumber \\
   &=& - g_0 (2j+1) \sum_\kappa A^\dagger_{001 \kappa} A_{001 \kappa}
        - G_2 \sum_{M \kappa} A^\dagger_{2M1 \kappa} A_{2M1 \kappa}
       \nonumber \\
   & & - X \sum_{J(\tau)} (-)^{J+1}5W(jjjj:2J)
     \sum_{M \kappa} A^\dagger_{JM \tau \kappa} A_{JM \tau \kappa},
        \label{eq:2} \\
 V^{\tau=0}_{\pi \nu} &=& - k^0 \sum_{J=odd}
       \sum_M A^\dagger_{JM00}  A_{JM00},  \label{eq:3}
\end{eqnarray}
with
\begin{displaymath}
 A^\dagger_{JM \tau \kappa} =  \sum_{m m^\prime}
 \langle j m j m^\prime |JM \rangle \sum_{\rho \rho^\prime}
 \langle \frac{1}{2} \rho \frac{1}{2} \rho^\prime |\tau \kappa \rangle
{1 \over \sqrt{2}} c^\dagger_{m \rho} c^\dagger_{m^\prime \rho^\prime},
\end{displaymath}
where $W(jjjj:2J)$ is the Racah coefficient. The interaction has four
parameters $g_0$, $G_2$, $X$ and $k^0$.

   Although the Coulomb energy cancels out in the double differences of
binding energies, it must be appropriately evaluated in order to
compare calculated energies of Hamiltonian (\ref{eq:1}) with observed binding
 energies.  We evaluate experimental energies of the ground states
 as follows:
\begin{equation}
 W_0(Z,N) = B(Z,N) - B(^{40}{\rm Ca}) - \lambda (A-40)
  - \Delta E_{C}(n_{p},n_{n}), \label{eq:4}
\end{equation}
where $B(Z,N)$ is the nuclear binding energy, $\lambda$ is the base
level of the single-particle state energy $\epsilon _{7/2}$ and 
$\Delta E_{C}(n_{p},n_{n})$ is
 the Coulomb energy correction for the valence nucleons
 $n=n_{p}+n_{n}$ ($n_{p}=Z-20$ and
 $n_{n}=N-20$).  Following Caurier {\it et al.} \cite{Caurier},
 we evaluate $\Delta E_{C}(n_{p},n_{n})$
 by the function
\begin{equation}
 \Delta E_{C}(n_{p},n_{n}) =
  7.279 \, n_{p} + 0.15 \, n_{p} (n_{p} -1)
  -0.065 \, n_{p} n_{n} . \label{eq:5}
\end{equation}
We fix $\lambda=-8.364$ MeV so that
 $W_0(^{41}{\rm Ca})=W_0(^{40}{\rm Ca})=0.0$, namely 
 $\epsilon_{7/2}=0.0$.
 
   We use the constant force strengths $g_0$=0.59, $G_2$=0.9 and
 $X$=1.2 in MeV neglecting $A$-dependence for the $P_0+P_2+QQ$ force
 and put $A$-dependence for the {\it p-n} force $V^{\tau=0}_{\pi \nu}$
 as $k^0=1.9\times(48/A)$ MeV.
 These parameters reproduce well the Coulomb corrected energies of
 valence nucleons near $^{48}$Cr \cite{Hase1}, and
 the double differences of binding energies for the $1f_{7/2}$ shell
 nuclei, especially the peak value at $N$=$Z$ \cite {Kane}.
 These results tell us that our model is reliable enough for our purpose
 to analyze respective contributions of interactions in the
 $\alpha$-like correlations.
 
   According to Ref. \cite{Hase2}, the ground states of 
 $(f_{7/2})^{4m}$ systems with
 $n_{p}$=$n_{n}$=$2m$ can be approximated by
 condensed states of the $I$=$T$=0 $\alpha$-like clusters
 (2p-2n units)
\begin{eqnarray}
& & |(f_{7/2})^{4m} {{\rm :gr}},I=T=0
    \rangle \approx {1 \over \sqrt{N_0}} (\alpha^\dagger_0)^m |A_0
    \rangle , \label{eq:6} \\
& & \alpha^\dagger_0 = \sum_{J \tau} \Psi^{(0)}(J \tau,J \tau:I=T=0)
    (A^\dagger_{J \tau} A^\dagger_{J \tau})_{I=0,T=0} , \nonumber
\end{eqnarray}
where $N_0$ is a normalization constant and $|A_0 \rangle$ is the
 doubly-closed-shell core state.  The approximate treatment of the 
 $(\alpha_0)^m$ model gives the energy $-32.04$ MeV to the exact
 energy $-32.20$ MeV for the $(f_{7/2})^8$ system.
  The modified treatment which takes the contributions of 2p-2n units
 with $I>0$ into account reproduces the energy $-32.20$ MeV.
 Therefore, the $I$=$T$=0 ground state of the $(f_{7/2})^8$ system
 is described as the dominant state $(\alpha_0)^2$ supplemented with
 other components of four-nucleon units ($\alpha_{I>0,T=0})^2$.
 (We showed in Ref. \cite{Hase2} that the $^{48}$Cr ground state is
 approximately described in terms of
 $\sum_I(\alpha_{I>0,T=0})^2$ in the full $fp$ space.)
  The squared amplitude of the dominant component $(\alpha_0)^2$ is
 0.973 in the present calculation.  The approximate treatment of
 the $(\alpha_0)^3$ model gives $-52.14$ MeV to the exact energy
 $-52.39$ MeV for the $(f_{7/2})^{12}$ system.  This result supports
 the goodness of our picture (\ref{eq:6}).
   The shell model calculations for the odd-odd systems
 $(f_{7/2})^{n=6,10}$ tell us that their ground states can
 be approximated by
\begin{equation}
 |(f_{7/2})^{4m+2}{{\rm :gr}},I=0,T=1
  \rangle \approx {1 \over \sqrt{N_1}} A^\dagger_{001\kappa}
 |(f_{7/2})^{4m}{{\rm :gr}},I=T=0 \rangle,
  \label{eq:7}
\end{equation}
where $N_1$ is a normalization constant. The overlap of the
 approximate state (\ref{eq:7}) with the exact state is 0.983 for the
 $(f_{7/2})^6$ system and 0.987 for the $(f_{7/2})^{10}$ system.
  The approximation (\ref{eq:7}) therefore holds well.
  We can say that the $\alpha$-like four-nucleon correlations play
 a dominant role in the $(f_{7/2})^{4m}$ systems with
 $n_{p}$=$n_{n}$=$2m$ and the $\tau$=1 monopole
 pairing correlation is also important when an additional
 proton-neutron pair exists.  Based on this knowledge,
 let us analyze competition of correlations.

   We calculated interaction energies of the $P_0+P_2+QQ$ force and
 $V^{\tau=0}_{\pi \nu}$ separately.  The calculated results for the
 isotopes with $n_{p}$=0, 2 and 4, which correspond to the Ca,
 Ti and Cr isotopes, are shown in Fig. 1.
 Look at the expectation value of the $P_0+P_2+QQ$ force for the
 ground states of the $n_{p}$=2 isotopes.
  As the neutron number $n_{n}$ increases, the energy gain
 increases more rapidly than that of the $n_{p}$=0 isotopes
 up to $n_{n}=2$ and the two lines of $n_{p}$=0 and
 $n_{p}$=2 become nearly parallel for $n_{n}\geq 2$.
   The line of $n_{p}$=4, which is parallel to that of
 $n_{p}$=2 up to $n_{n}$=2, shows an additional
 energy gain at $n_{n}$=$n_{p}$=4 and becomes again
 parallel to that of $n_{p}$=2 for $n_{p} \geq 4$.
   We know that a considerably large energy gain of the $P_0+P_2+QQ$
 force is attained every time when a 2p-2n unit with $T$=0 is formed.
   At the points $n_{n}$=$n_{p}$=2 and 4, in
 Fig. 1, the energy gain of $V^{\tau=0}_{\pi \nu}$ also shows a peak
 (see the bend in the line of $\langle V^{\tau=0}_{\pi \nu} \rangle$).
  The ground states of the $(f_{7/2})^{4m}$ systems with
 $n_{n}$=$n_{p}$=$2m$ have the $\alpha$-like
 correlated structure as mentioned above.  Therefore, the large energy
 gain of the $P_0+P_2+QQ$ force at
 $n_{n}$=$n_{p}$=$2m$ drives the system to form the
 $\alpha$-like units.  It should be remembered here that the 
 ground-state wavefunction is determined by the $P_0+P_2+QQ$ force
 \cite{Hase1}.
   The {\it p-n} interaction $V^{\tau=0}_{\pi \nu}$ which is written as
 $-\frac{1}{2} k^0 \{ {\hat n \over 2} ({\hat n \over 2} + 1)
 - {\hat T^2} \}$ endows the $T=0$ $\alpha$-like correlated state
 with a very large energy gain.
 
  To discuss the large binding energies of the
 $n_{n}$=$n_{p}$=$2m$ systems due to the
 $\alpha$-like four-nucleon correlations, Gambhir, Ring and Schuck
 \cite{Gambhir} considered differences of binding energies
\begin{eqnarray}
 S(A_0+4m+2)&=& B(Z_0+2m, N_0+2m) \nonumber \\
      &-& {1 \over 2} \{B(Z_0+2m+2,N_0+2m) +B(Z_0+2m,N_0+2m+2)\},
  \label{eq:8} \\
 S(A_0+4m+4)&=& {1 \over 2} \{B(Z_0+2m+2,N_0+2m)+B(Z_0+2m,N_0+2m+2)\}
           \nonumber \\
      &-& B(Z_0+2m+2,N_0+2m+2).      \label{eq:9}
\end{eqnarray}
They explained in terms of the $\alpha$-like correlated units that
 $S(A_0+4m+4)$ is systematically larger than $S(A_0+4m+2)$.  The two
 quantities $S(A_0+4m+4)$ and $S(A_0+4m+2)$ have contributions from
 the Coulomb energy.
   If we consider the difference $S(A_0+4m+4)-S(A_0+4m+2)$, the
 contributions from the Coulomb energy and single-particle energy
 nearly disappear (though a small effect of the Coulomb energy remains
 as long as we adhere the approximation (\ref{eq:5})).
  It is convenient to take up the Coulomb corrected energy of valence
 nucleons $W_0(Z,N)$ instead of $B(Z,N)$, because the effective
 Hamiltonian aims to reproduce $W_0(Z,N)$.  We are considering the 
 $N \approx Z$ nuclei in which protons and neutrons occupy the same
 shell.  The ground states of the $A_0+4m+2$ systems are the $I$=0,
 $T$=1 states (in the $f_{7/2}$ shell nuclei) and the Hamiltonian is
 isospin invariant.  Hence, we have the approximate relations
\begin{eqnarray}
 W_0(Z_0+2m+2,N_0+2m) & \approx & W_0(Z_0+2m, N_0+2m+2) \nonumber \\
       & \approx & W_0(Z_0+2m+1,N_0+2m+1).  \label{eq:10}
\end{eqnarray}

   Let us define the following quantities corresponding to Eqs.
 (\ref{eq:8}) and (\ref{eq:9}):
\begin{eqnarray}
  S^\prime(n_{p}+2,n_{n})
   &=& W_0(n_{p},n_{n})
    - W_0(n_{p}+2,n_{n}) ,   \label{eq:11} \\
  S^\prime(n_{p}+2,n_{n}+2)
   &=& W_0(n_{p},n_{n}+2)
    - W_0(n_{p}+2,n_{n}+2) ,   \label{eq:12}
\end{eqnarray}
 When neglecting the small Coulomb energy effect, we can expect
\begin{eqnarray}
  \delta S^\prime(n_{p}+2,n_{n}+2) &=&
  S^\prime(n_{p}+2,n_{n}+2)
  -S^\prime(n_{p}+2,n_{n})  \nonumber \\
  & \approx & S(A_0+4m+4) - S(A_0+4m+2).  \label{eq:7-2}
\end{eqnarray}
  The difference $\delta S^\prime(n_{p}+2,n_{n}+2)$
 =$S^\prime(n_{p}+2, n_{n}+2)
 -S^\prime(n_{p}+2, n_{n})$ represents the
 correlations between two proton and two neutrons from the 
 viewpoint of Gambhir, Ring 
 and Schuck\cite{Gambhir}.  The same quantity (which is called the
 double difference of binding energies) has recently been investigated
 as the {\it p-n} interaction in Refs. \cite{Zhang,Brenner}, where the
 remarkable increase of this quantity at
 $N$=$Z$ (or $n_{n}=n_{p}$) is also discussed.

  The quantity $S^\prime(n_{p}+2,n)$ corresponds to
 $(A_n-B_n)+(a_n-b_n)$ and $(B_n-C_n)+(b_n-c_n)$ in Fig. 1.
   For instance, $S^\prime(4, 4)$ is shown by the two solid lines
 connecting $B_n$ and $C_n$ ($b_n$ and $c_n$) at $n_{n}$=4,
 and $S^\prime(4, 2)$ is shown by the two broken lines at
 $n_{n}$=2.  Similarly, $S^\prime(2, 2)$ and $S^\prime(2, 0)$
 are shown by the solid and broken lines connecting $A_n$ and $B_n$
 ($a_n$ and $b_n$) at $n_{n}$=2 and 0, respectively.
   The solid line representing $S^\prime(2m+2, 2m+2)$ is longer than
 the broken line representing $S^\prime(2m+2, 2m)$ both in
 $\langle P_0+P_2+QQ \rangle$ and
 $\langle V^{\tau=0}_{\pi \nu}\rangle$.
   This was discussed as the evidence of the $\alpha$-like four-nucleon
 correlations by Gambhir, Ring and Schuck \cite{Gambhir}.
 They, however, did not consider the {\it p-n} interaction like
 $V^{\tau=0}_{\pi \nu}$ and hence could not reproduce sufficient values
 for $S(A_0+4m+4)-S(A_0+4m+2)$.  We know the indispensable role of
 $V^{\tau=0}_{\pi \nu}$ in the $\alpha$-like four-nucleon correlations.
 
   Thus, the double difference
 $\delta S^\prime(n_{p}+2,n_{n}+2)$ is a good measure
 of the $\alpha$-like correlation energy.  We showed in Ref. \cite{Kane}
 that $\delta S^\prime$ has a triangular-shape peak at $N=Z$.
 This can be visually understood in Fig. 1 as follows:
 The differences
  $(b_n-c_n)-(b_{n-2}-c_{n-2})$ and $(B_n -C_n)-(B_{n-2}-C_{n-2})$
 have maximum values at $n_{n}$=4 and are secondly large at
 $n_{n}$=3 and 5.
 The quantity $(B_n-C_n)-(B_{n-2}-C_{n-2})$ becomes negligible when
 $|n_{n}-n_{p}|>1$, since the two lines of
 $n_{p}$=2 and 4 for $\langle P_0+P_2+QQ \rangle$ are
 nearly parallel when $n_{n} \leq 2$ and
 $n_{n} \geq 4$, namely $\delta S^\prime$ approximately
 comes from $V_{\pi \nu}^{\tau =0}$ and becomes roughly a constant
 $k^0/2$ for $|n_{n}-n_{p}|>1$.

   To analyze the interesting behavior of the {\it p-n} interaction
 (correlations between two protons and two neutrons) in detail, 
 let us now see the
 contributions of {\it p-n} and like-nucleon ({\it p-p} and 
 {\it n-n}) correlations
 separately.  Since we know the behavior of the {\it p-n} interaction
 $V^{\tau =0}_{\pi \nu}$, we show the two types of contributions
 from the $P_0+P_2+QQ$ force in Fig. 2.
 This figure demonstrates that the p-n interaction of the
 $P_0+P_2+QQ$ force gives energy gains roughly proportional to the
 neutron number till $n_{n}$=$n_{p}$, and attains
 a maximum energy gain when a maximum number of $\alpha$-like units
 are formed at $n_{n}$=$n_{p}$=$2m$.
 Once all the protons join the $\alpha$-like units, additional neutrons
 interact weakly with the protons in the $\alpha$-like units through
 the {\it p-n} interaction of the $P_0+P_2+QQ$ force (somewhat decreasing
 the energy gain attained by forming the $\alpha$-like units).
 On the other hand, the {\it n-n} correlations between the additional
 neutrons are still observed in Fig. 2.
 
    Accordingly, we reach the following conclusion. The remarkable peak
 observed at $N=Z$ in the double difference of binding energies is
 closely related to the $\alpha$-like four-nucleon correlations.
 The maximum contribution of the {\it p-n} interaction in cooperation with
 the like-nucleon interactions within the $P_0+P_2+QQ$ force,
 forming the $\alpha$-like correlated structure,
 attains a spetial energy gain in the $N$=$Z$ even-even systems.
 Another {\it p-n} interaction $V^{\tau =0}_{\pi \nu}$ makes the structure
 stable by endowing a very large energy gain.  It is interesting that
 the coupling between the $\alpha$-like units and residual neutrons
 is weak in the $P_0+P_2+QQ$ force.  This suggests a direct product
 state of the $\alpha$-like correlated structure and residual pairing
 correlated neutrons at the end.  In the $N$=$Z$ odd-odd systems,
 the $\tau$=1 monopole pairing interaction is important for the last
 {\it p-n} pair and the wavefunction is approximated by Eq. (\ref{eq:7}).
   We have made our study within the $(f_{7/2})^n$ model, in this
 note.  The collective feature of the $\alpha$-like four-nucleon
 correlations over many $j$ shells could affect the binding energy
 little but may considerably change the detailed structure.
 Still, the outline of our interpretation about the p-n interactions,
 the double difference of binding energies and the $\alpha$-like
 correlations using the $P_0+P_2+QQ+V_{\pi \nu}^{\tau =0}$ interaction
 will remain valid.
 The $\alpha$-like correlations should be stressed even in the shell
 model approach to the $fp$ shell nuclei.


\begin{figure}
\begin{center}
 \leavevmode
 \epsfxsize=12cm
 \epsfysize=18cm
 \epsfbox{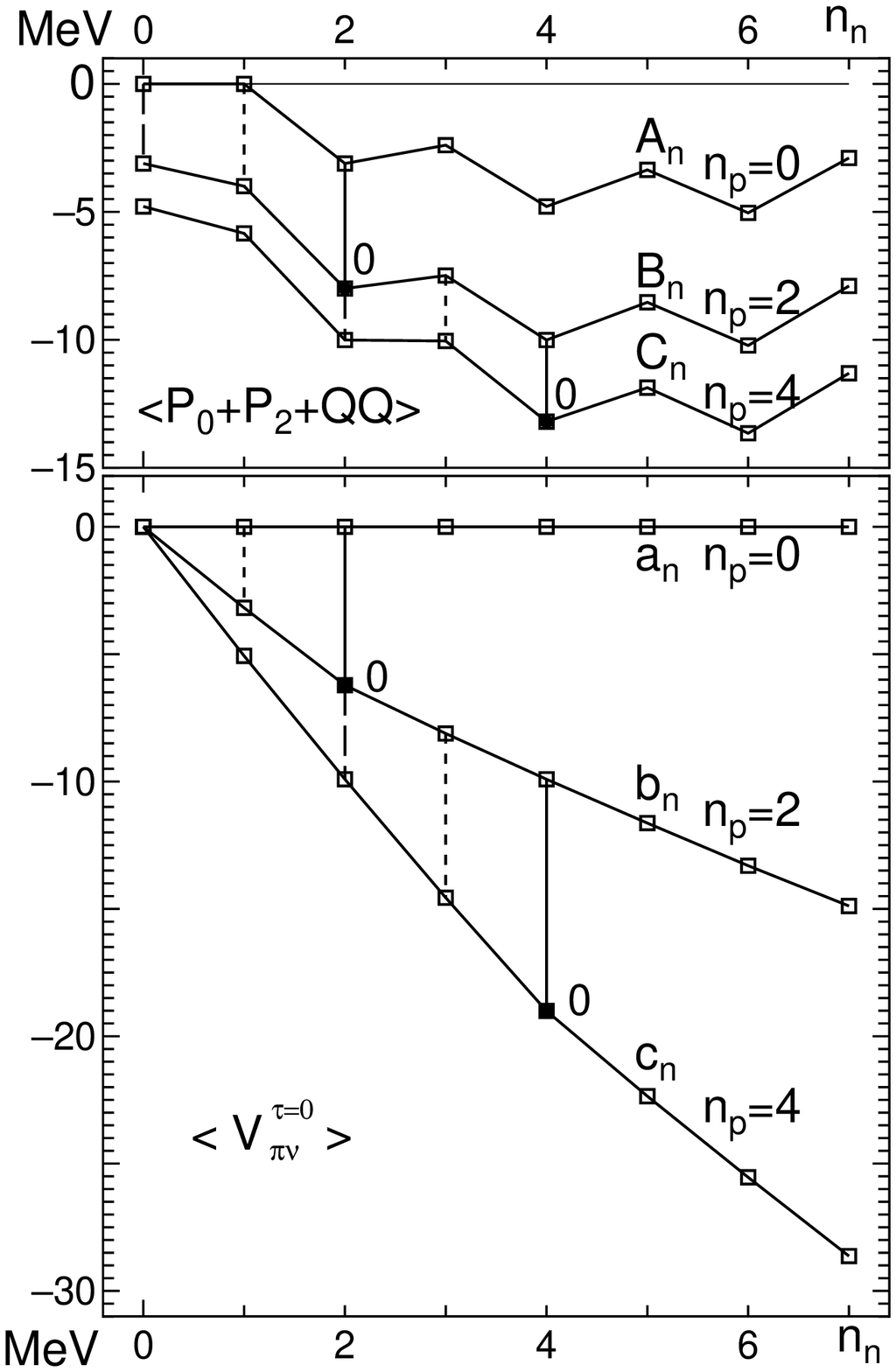}
\caption{Interaction energies of the $P_0+P_2+QQ$ force and
 $V^{\tau =0}_{\pi \nu}$ in the $n_{p}$=0, 2 and 4 isotopes.
 The states with $T$=0 are shown by the solid squares.
 The isospin increases as aparting from the solid squares.}
\label{autonum}
\end{center}
\end{figure}

\newpage
\begin{figure}
\begin{center}
 \leavevmode
 \epsfxsize=12cm
 \epsfysize=14cm
 \epsfbox{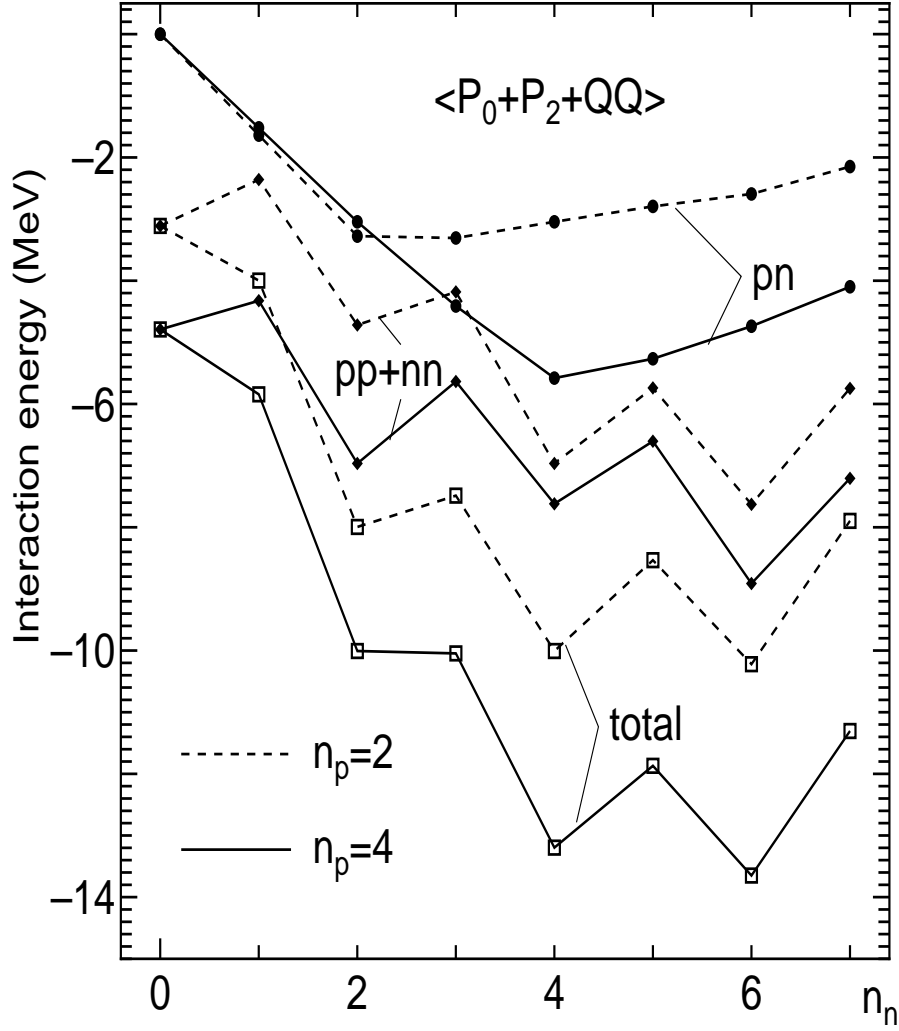}
\caption{{\it p-n} and {\it p-p} plus {\it n-n} components 
of interaction energy of
 the $P_0+P_2+QQ$ force in the $n_{p}$=2 and 4 isotopes.}
\end{center}
\end{figure}

\end{document}